\documentclass{optica-article}

\journal{opticajournal} 

\articletype{Research Article}

\usepackage{lineno}

\begin{document}

\title{Galvanometer Mirror Based Femtosecond-Laser Volumetric Displays}

\author{Katsuhito Muroi,\authormark{1,*} Ryosei Kojima,\authormark{2} Tatsuki Fushimi\authormark{1,3,4} and Yoichi Ochiai,\authormark{1,3,4,5}}

\address{\authormark{1} R\&D Center for Digital Nature, University of Tsukuba, Tsukuba, 305-8550, Ibaraki, Japan\\
\authormark{2} School of Informatics, College of Media Arts, Science and Technology, University of Tsukuba, Kasuga Campus Kasuga 1-2, Tsukuba, Ibaraki, 305-8550, Japan\\
\authormark{3} Institute of Library, Information and Media Science, University of Tsukuba, Tsukuba, 305-8550, Ibaraki, Japan\\
\authormark{4} Tsukuba Institute for Advanced Research (TIAR), University of Tsukuba, 1-1-1 Tennodai, Tsukuba, Ibaraki, 305-8577, Japan\\
\authormark{5} Pixie Dust Technologies, Inc., Chuo-ku, 104-0028, Tokyo, Japan}

\email{\authormark{*}katsuhitomuroi@digitalnature.slis.tsukuba.ac.jp} 


\begin{abstract}{
Femtosecond-laser-induced plasma enables self-emissive voxels to be generated directly in air for drawing three-dimensional images. Galvanometer-mirror-based beam scanning provides a method for rapidly moving plasma voxels using a relatively simple optical configuration. However, the generated plasma voxels are not ideal points but luminous regions with finite lengths and widths. Variations in voxel shape and orientation caused by focusing and scanning conditions may therefore affect the appearance of rendered points and lines. Nevertheless, quantitative relationships between focusing and scanning conditions and plasma-voxel geometry remain insufficiently characterized in systems using a two-axis galvanometer mirror and a single focusing lens. Here, we quantitatively investigate how focal length and two-axis galvanometer scan angle affect the shape of plasma voxels and the projected inclination of their principal axes. For focal lengths ranging from 75 to 300~mm, the mean plasma length increased from 1.39 to $4.64~\mathrm{mm}$. These results indicate that increasing the focal length produces more elongated and anisotropic plasma voxels. The projected inclination of the plasma principal axis varied approximately linearly with the corresponding scan angle, although the response differed between the two scanning axes. These results provide quantitative design data for selecting focusing conditions and correcting voxel shape and orientation according to scanning conditions in galvanometer-mirror-based aerial volumetric displays.}
\end{abstract}

\section{Introduction}

Volumetric displays form emissive or light-scattering points in three-dimensional space and can present three-dimensional images with motion parallax over multiple viewing directions without requiring the observer to wear special glasses \cite{Geng2013,Holliman2011}. One approach is to form self-emissive voxels directly in air using laser-induced plasma \cite{Saito2008,Ochiai2016}. In particular, femtosecond-laser-based systems can generate visible luminous points with a lower single-pulse energy than nanosecond-laser-based systems, enabling aerial displays without screens or scattering media \cite{Ochiai2016}. Previous studies have investigated multifocal control using computer-generated holograms, tactile presentation, expansion of the drawing region, multicolor operation, and enlargement of the display volume using multiple optical paths \cite{Ochiai2016,Kumagai2021,Kumagai2025,Kumagai2017,Hayasaki2018}.

The position of a plasma voxel is determined by the high-intensity focal region of the excitation laser beam. Holographic wavefront control using a spatial light modulator can flexibly generate one or multiple focal points in three-dimensional space and can also be used for parallel generation of multiple voxels \cite{Ochiai2016,Kumagai2021,Hayasaki2018}. However, this approach requires computer-generated phase patterns and additional optical components, while the available optical power, diffraction efficiency, and refresh rate depend on the characteristics of the modulator. In contrast, a two-axis galvanometer mirror system with a single focusing lens enables high-speed angular scanning of the laser beam \cite{Saito2008,Ochiai2016,Kumagai2025}. In such a scanning system, the focusing conditions vary with focal length and scan angle, and the plasma-voxel shape is therefore expected to vary accordingly. The focusing conditions determined by the external optics have been reported to strongly influence plasma density, filament diameter, and the optical-breakdown region during femtosecond-laser propagation in air \cite{Braun1995,Couairon2007,Berge2007,Kasparian2008,Chin2012,Theberge2006,Liu2010}. In addition, pre-objective two-axis galvanometer mirror scanning can produce angle-dependent geometric distortion and asymmetry \cite{Verboven1988}. Consequently, voxel length, diameter, and inclination may vary across the display region, causing the appearance of rendered points and lines to depend on display position and observation angle. Although scan-optimized optical systems can reduce these effects, quantitative data directly relating focal length and two-axis scan angle to measured voxel shape and orientation remain limited.

In this study, we quantitatively evaluate the effects of focal length and scan angle on the shape and orientation of plasma voxels in a femtosecond-laser plasma volumetric display consisting of a two-axis galvanometer mirror and interchangeable single focusing lenses. Plasma length, diameter, and aspect ratio are measured for different focal lengths, while the projected inclination of the plasma principal axis is obtained from top-view and side-view images for two-axis scan angles. This analysis quantifies changes in plasma-voxel shape and projected orientation in a basic single-lens configuration and examines the relationship between display range and voxel shape.

\section{Methods}
\subsection{Experimental optical system and coordinate definition}
\begin{figure}
    \centering
    \includegraphics[width=0.85\linewidth]{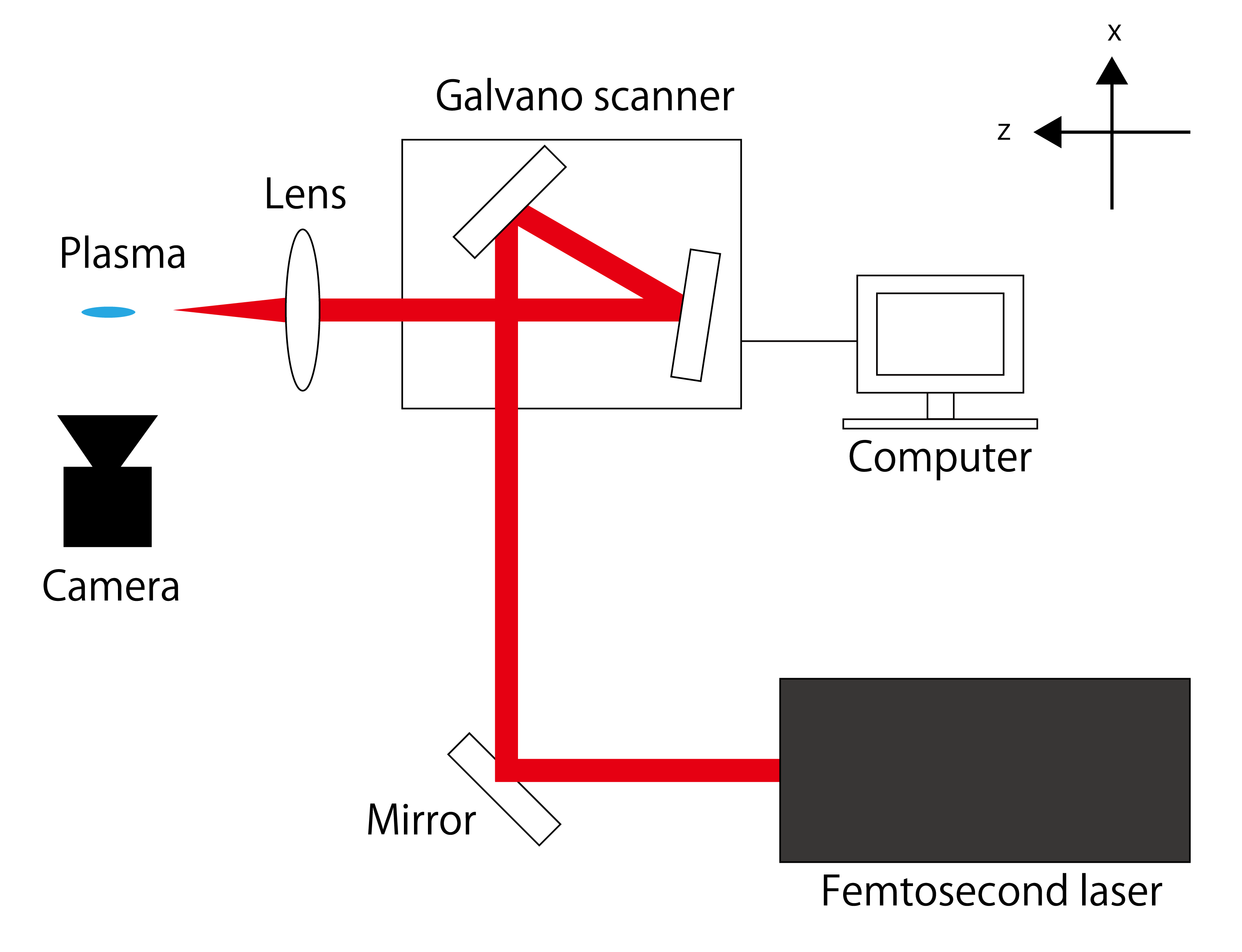}
    \caption{Experimental setup}
    \label{fig:placeholder}
\end{figure}
A Ti:sapphire femtosecond laser system (Astrella USP, Coherent) was used. The center wavelength was $800~\mathrm{nm}$, the repetition rate was $1~\mathrm{kHz}$, and the measured pulse duration was $41~\mathrm{fs}$. The average power immediately before the focusing lens, after the galvanometer mirror, was measured to be $3.7~\mathrm{W}$ using a FieldMate power meter with a PM10 thermopile sensor (Coherent). The manufacturer's nominal $1/e^2$ intensity beam diameter of $11~\mathrm{mm}$ was used in the theoretical calculation.

The laser beam was directed to a two-axis galvanometer mirror and, after deflection, focused in air by a single plano-convex lens. For the focal-length-dependence measurements, plano-convex lenses with focal lengths of $f = 75$, 100, 125, 150, 200, and $300~\mathrm{mm}$ were used interchangeably. Plasma emission was generated at the focus by optical breakdown of air induced by the femtosecond laser pulses.

The beam was deflected by a two-axis galvanometer mirror scanner configured with large-aperture fused-silica broadband dielectric mirrors (Thorlabs BBSQ1-E03, $25.4 \times 25.4 \times 6.0~\mathrm{mm}$, for the $X$ axis, and a BBSQ2-E03 mirror bisected to $50.8 \times 25.4 \times 6.0~\mathrm{mm}$ for the $Y$ axis). These mirrors carry an E03 dielectric coating ($R_{\mathrm{avg}} > 99\%$ over $750$--$1100~\mathrm{nm}$ at $0^\circ$--$45^\circ$ incidence) on fused-silica substrates specified to $\lambda/10$ surface flatness at $633~\mathrm{nm}$ and $10$--$5$ surface quality; the values quoted are those of the uncut substrates. The rectangular $Y$-axis mirror accommodates the lateral beam displacement introduced by the $X$-axis mirror in addition to the $11 \times 15.6~\mathrm{mm}$ elliptical footprint formed by the $\varnothing 11~\mathrm{mm}$ beam at $45^\circ$ incidence. The specified clear aperture ($>85\%$ of the $25.4~\mathrm{mm}$ dimension) transmits $99.4\%$ of this beam footprint, and the combined wavefront error of the two mirrors corresponds to a Strehl ratio above 0.9. The peak fluence on the scanner mirrors was $5.5~\mathrm{mJ/cm^2}$ (peak intensity $1.3 \times 10^{11}~\mathrm{W/cm^2}$), a factor of 37 below the specified coating damage threshold of $0.205~\mathrm{J/cm^2}$ measured at $800~\mathrm{nm}$, $99~\mathrm{fs}$, and $1~\mathrm{kHz}$.

The galvanometer mirror scan angles were calibrated in advance using the displacement of the laser spot on a screen. Because geometric distortion can arise from the mirror arrangement and scan angle in a two-axis galvanometer mirror system \cite{Verboven1988}, the control value was not directly treated as an angle. Instead, the deflection angle was obtained from the beam displacement in free space. During calibration, the screen was moved to several axial positions while a fixed control value was applied to the galvanometer mirrors, and the beam position was recorded at each screen position. The beam deflection angle was derived from the relationship between screen distance and spot displacement, thereby establishing the correspondence between galvanometer mirror control value and scan angle. This calibration was incorporated into the control program, which then supplied the galvanometer mirror with the control value corresponding to the specified angle.

The central optical axis of the lens was defined as the $Z$ axis, and the two scan directions of the galvanometer mirror were defined as the $X$ and $Y$ axes. The calibrated scan angles in the $X$ and $Y$ directions were denoted by $\theta_x$ and $\theta_y$, respectively. The projected inclination angles of the plasma principal axis obtained from the top-view and side-view images were denoted by $\phi_x$ and $\phi_y$, respectively. This calibration determined the correspondence between the galvanometer mirror control value and the free-space beam-deflection angle; it did not directly measure the actual incidence angle at the focusing lens. Therefore, the scan angles reported in this study are treated as calibrated values based on prior beam-displacement measurements.

\subsection{Measurement conditions}
\subsubsection{Focal-length-dependence measurements}

To evaluate the change in plasma-voxel shape with focal length, plano-convex lenses with focal lengths of $f = 75$, 100, 125, 150, 200, and $300~\mathrm{mm}$ were used. The lenses were selected from a $2$-inch N-BK7 plano-convex lens kit with an antireflection coating for $650$-$1050~\mathrm{nm}$ (LSC01-B, Thorlabs). During the measurements, both galvanometer mirror scan angles were fixed at $0^\circ$, and the average laser power immediately before the focusing lens was maintained at $3.7~\mathrm{W}$.

For each focal length, the camera settings were a shutter speed of $1/250$~s, an aperture of $F/5.6$, and an ISO sensitivity of 2500. For focal lengths from 75 to $150~\mathrm{mm}$, a neutral-density filter with 50\% transmittance was used to suppress scattered light. No neutral-density filter was used for focal lengths of 200 and $300~\mathrm{mm}$.

As the plasma-generation position shifted with focal length, the camera was translated along the $Z$ axis to keep the plasma image centered. The camera height, distance from the plasma, and imaging magnification were kept constant. The spatial scale of each image was determined from a calibration image acquired by placing a $1~\mathrm{cm} \times 1~\mathrm{cm}$ grid at the plasma-generation position. Plasma length, diameter, and aspect ratio were then evaluated from the acquired images.

To evaluate the inclination of the plasma-emission region as a function of galvanometer mirror scan angle, the focal length was fixed at $f = 100~\mathrm{mm}$, and the scan angles in the $X$ and $Y$ directions were varied. For each axis, five scan angles were used: $-7.0^\circ$, $-3.5^\circ$, $0^\circ$, $3.5^\circ$, and $7.0^\circ$. All combinations of the $X$- and $Y$-direction scan angles were measured, resulting in a total of 25 conditions.

For each scan condition, the plasma-emission region was recorded from both the side and the top. The side-view images were used primarily to evaluate the component of plasma inclination associated with the $Y$ direction, whereas the top-view images were used primarily to evaluate the component associated with the $X$ direction. The distance between the camera and the plasma-emission region was approximately $300~\mathrm{mm}$. The camera settings were a shutter speed of $1/60$~s, an aperture of $F/8$, and an ISO sensitivity of 5000. No optical filter was used.

One side-view image and one top-view image were acquired for each scan condition. No background subtraction was performed. The plasma-emission region was extracted from each image, and its principal-axis direction was determined by image analysis. The plasma inclination angles obtained from the top-view and side-view images were denoted by $\phi_x$ and $\phi_y$, respectively.

Because only one image was acquired under each condition, error bars based on image-to-image variation were not calculated. The relationship between scan angle and projected inclination angle was instead evaluated using the slope, intercept, and coefficient of determination of a linear fit.

\subsection{Image analysis}
\subsubsection{Plasma length and diameter}

Plasma length and width were evaluated from side-view JPEG images acquired at each focal length. For each pixel, the maximum value among the red, green, and blue channels was used as the intensity signal. An initial reference region was identified using a threshold of 20\% of the maximum pixel value within the region of interest (ROI). The peak pixel value within this reference region was then used to determine a relative threshold of 35\% for extracting the plasma-emission region. For the width measurement at $f = 75~\mathrm{mm}$, a threshold of 45\% was used to reduce the influence of the surrounding diffuse glow. The effect of threshold selection on the measured plasma width is presented in the Supplementary Document.

After thresholding, a morphological opening operation with a $3\times3$ elliptical structuring element was applied, and the connected component with the largest overlap with the reference region was selected. Plasma length was defined as the horizontal extent of the extracted emission region, whereas plasma width was defined as its maximum vertical extent in the side-view image. The measured pixel dimensions were converted into millimeters using the corresponding horizontal and vertical calibration factors determined from calibration images acquired at each focal length.

For each focal length, ten images were acquired under identical experimental and imaging conditions. The mean and sample standard deviation of the plasma length and width were calculated from the ten measurements.

\subsubsection{Projected inclination angle of the plasma principal axis}

The projected inclination angle of the plasma principal axis was obtained from the top-view and side-view images acquired in the scan-angle-dependence measurements. For each pixel, the maximum value among the red, green, and blue channels was used as the intensity signal, and Gaussian smoothing with a standard deviation of 1.2 pixels was applied to the resulting RGB-maximum-intensity image.

The emission region was extracted by thresholding the smoothed image. The threshold was set to 18\% of the peak intensity in each image. To prevent background noise from being extracted as an emission region in low-intensity images, the threshold was constrained to a minimum value of 12. Thus, the threshold $T$ applied to each image was
\begin{equation}
    T = \max\left(0.18 I_{\max}, 12\right),
    \label{eq:threshold}
\end{equation}
where $I_{\max}$ is the maximum intensity in the smoothed image.

After binarization, small isolated noise components were removed using a morphological opening operation with an elliptical $3\times3$-pixel kernel. A morphological closing operation with an elliptical $7\times3$-pixel kernel was then used to fill small gaps and discontinuities within the plasma-emission region.

The connected component containing the maximum-intensity pixel was selected as the emission region. If the area of this component was less than 20 pixels, the largest connected component was selected instead. This fallback procedure was not invoked for any of the 50 images analyzed in this study.

Principal component analysis (PCA) was applied to the coordinates of all pixels in the selected emission region, and the direction of maximum spatial variance was obtained as the first principal component \cite{Jolliffe2016}. This first principal component was defined as the plasma principal axis, and its angle relative to the horizontal direction of the image was defined as the projected inclination angle.

The projected inclination angle in the $XZ$ plane obtained from the top-view images was denoted by $\phi_x$, and that in the $YZ$ plane obtained from the side-view images was denoted by $\phi_y$. For all 25 combinations of the $X$- and $Y$-direction scan angles, $\phi_x$ was calculated from each top-view image and $\phi_y$ from each side-view image. These angles were treated as projections of the plasma principal axis onto the respective imaging planes rather than as a combined three-dimensional inclination angle.

\section{Results}
\subsection{Variation in plasma-voxel shape with focal length}
\begin{figure}
    \centering
    \includegraphics[width=0.7 \linewidth]{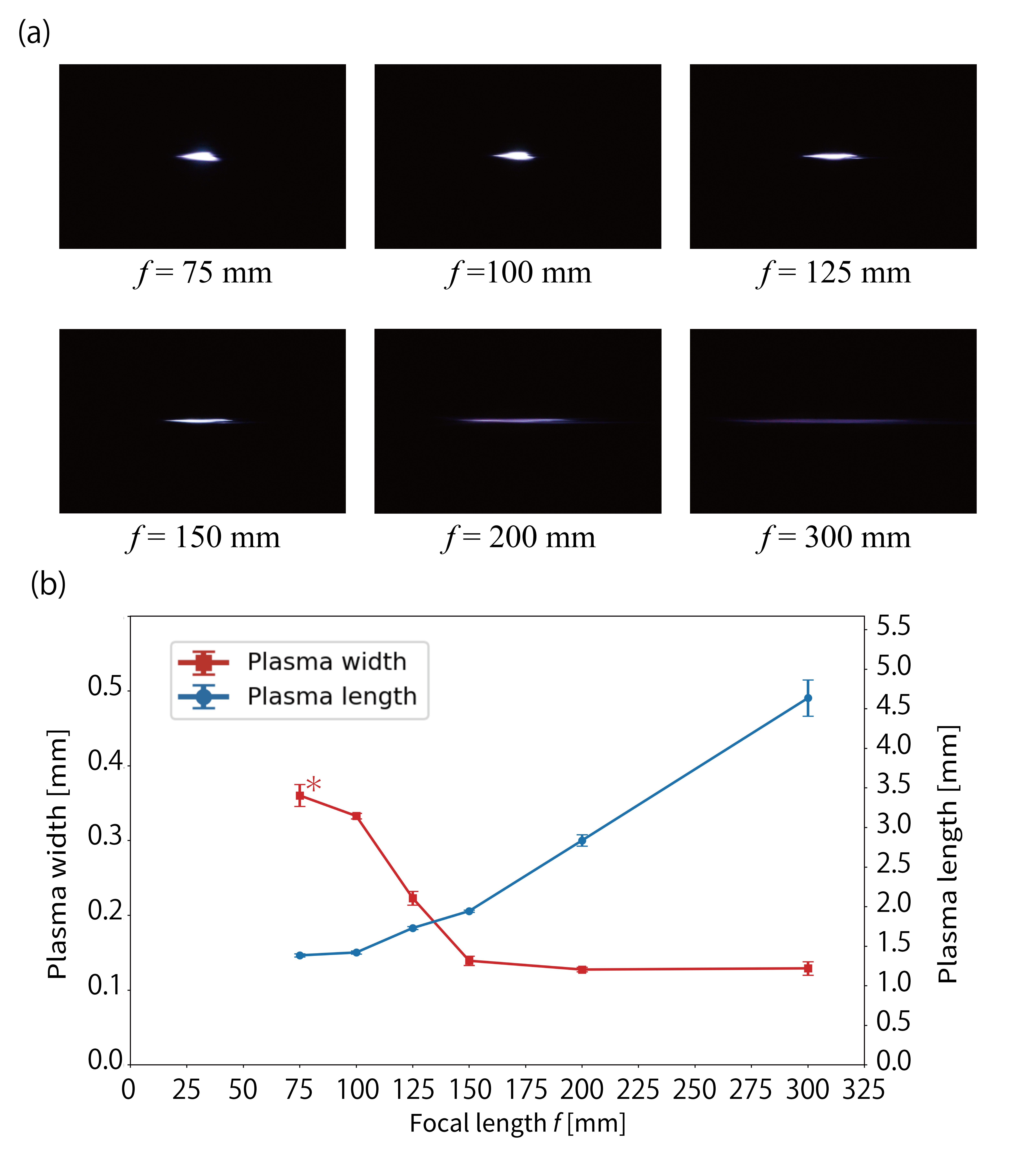} 

\caption{Effect of focal length on the dimensions of the laser-induced plasma. (a) Representative side-view images of the plasma generated using focusing lenses with focal lengths ranging from 75 to 300~mm. (b) Measured plasma length and width as functions of focal length. Data points represent the mean values of ten measurements, and error bars indicate one standard deviation ($n = 10$). Plasma length and width were measured using a relative intensity threshold of 35\%, except for the plasma width at $f = 75~\mathrm{mm}$ (marked with *), which was measured using a threshold of 45\%. See Supplementary Section 3 and Fig.~S2 for details.}
    \label{fig:placeholder}
\end{figure}

Figure~2(a) shows representative side-view images of the plasma emission acquired for focal lengths from 75 to 300~mm, while Fig.~2(b) shows the measured plasma length and width as functions of focal length. As the focal length increased, the plasma-emission region extended along the optical axis. The mean plasma lengths were 1.39~mm and 4.64~mm at $f = 75~\mathrm{mm}$ and $300~\mathrm{mm}$, respectively. The plasma length increased gradually from 75 to 150~mm, followed by a pronounced increase between 150 and 200~mm. A further increase was observed between 200 and 300~mm. The mean plasma widths were 0.36~mm and 0.14~mm at $f = 75~\mathrm{mm}$ and $150~\mathrm{mm}$, respectively. At focal lengths of 200 and 300~mm, the plasma width remained nearly constant at approximately 0.13~mm. These results show that increasing the focal length produced a plasma voxel that was increasingly elongated along the optical axis.

\subsection{Projected inclination of the plasma principal axis as a function of scan angle}

\begin{figure}
    \centering
    \includegraphics[width=0.71\linewidth]{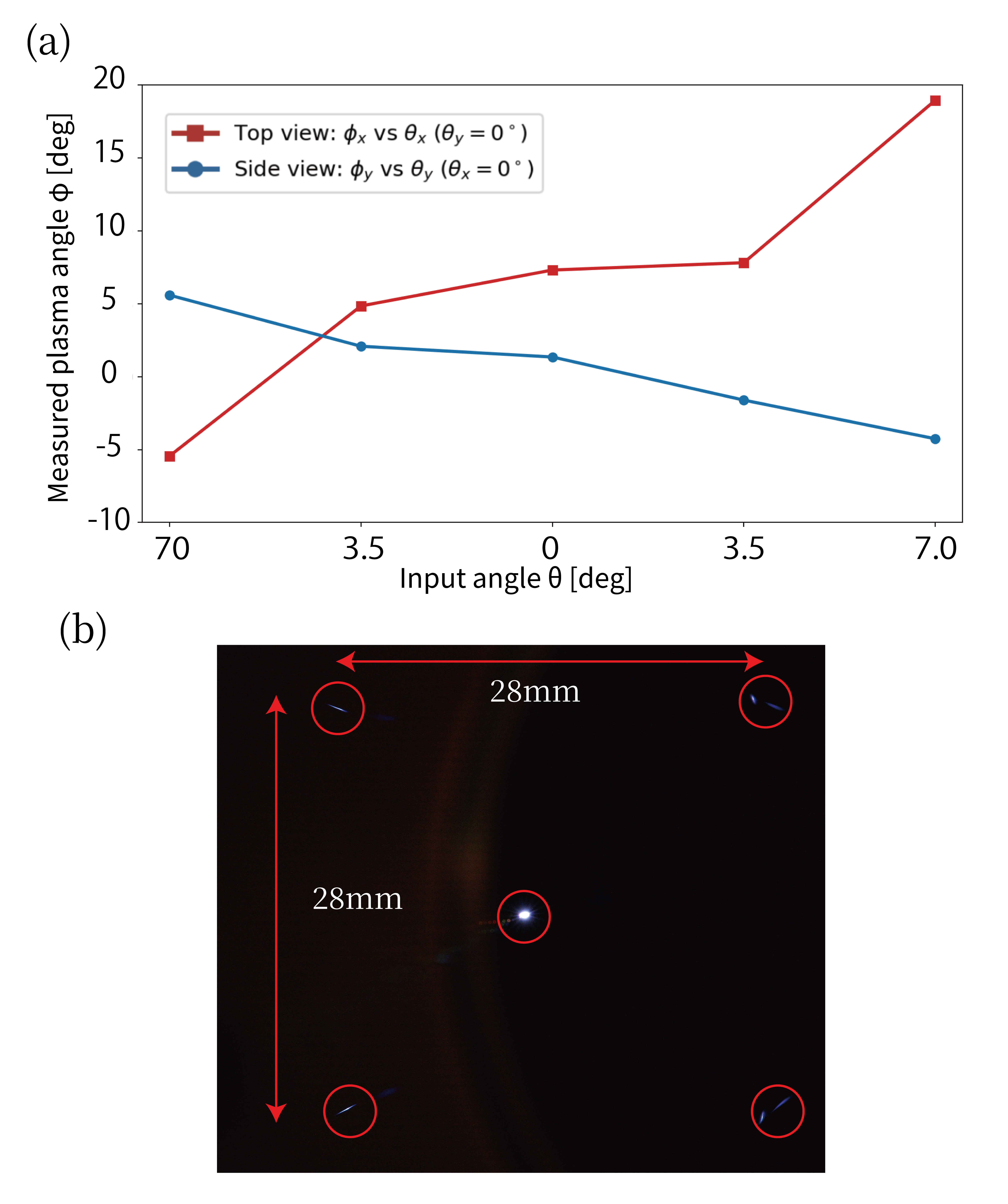}
    \caption{Plasma orientation and generation positions at $f = 100~\mathrm{mm}$. 
(a) Plasma angle $\phi$ as a function of input angle $\theta$. 
Red: $\phi_x$ versus $\theta_x$ obtained from top-view images at $\theta_y = 0^\circ$. 
Blue: $\phi_y$ versus $\theta_y$ obtained from side-view images at $\theta_x = 0^\circ$. 
(b) Composite image constructed from separately acquired plasma images recorded from the same camera position for $(\theta_x,\theta_y) = (\pm 7^\circ,\pm 7^\circ)$ and $(0^\circ,0^\circ)$, spanning approximately $28 \times 28~\mathrm{mm}^2$.}
    \label{fig:placeholder}
\end{figure}

Figure~3 shows the plasma inclination angles as functions of the $X$- and $Y$-direction scan angles. The angle $\phi_x$ obtained from the top-view images depended mainly on the $X$-direction scan angle $\theta_x$, whereas the effect of $\theta_y$ at a fixed $\theta_x$ was relatively small. Similarly, $\phi_y$ obtained from the side-view images depended mainly on the $Y$-direction scan angle $\theta_y$. With $\theta_y=0^\circ$, $\phi_x$ was $-5.44^\circ$ and $18.96^\circ$ at $\theta_x=-7^\circ$ and $7^\circ$, respectively. Linear regression yielded $\phi_x=1.48\theta_x+6.70$, with a coefficient of determination of $R^2=0.89$. With $\theta_x=0^\circ$, $\phi_y$ was $5.59^\circ$ and $-4.26^\circ$ at $\theta_y=-7^\circ$ and $7^\circ$, respectively. Linear regression yielded $\phi_y=-0.67\theta_y+0.64$, with a coefficient of determination of $R^2=0.97$.
These results confirm that the projected inclination angle of the plasma principal axis varied approximately linearly with the corresponding galvanometer scan angle. Across all 25 conditions, some variation associated with the orthogonal scan axis was also observed; nevertheless, $\phi_x$ was predominantly dependent on $\theta_x$, and $\phi_y$ was predominantly dependent on $\theta_y$.

The physical plasma-generation positions corresponding to the scan-angle range were also examined. Plasma was generated at the four conditions $(\theta_x,\theta_y)=(7^\circ,7^\circ)$, $(-7^\circ,7^\circ)$, $(-7^\circ,-7^\circ)$, and $(7^\circ,-7^\circ)$. At the axial position of the on-axis plasma, the four laser irradiation marks formed an approximately $28 \times 28~\mathrm{mm}^2$ square on an acrylic plate. At the frontmost position of the off-axis plasma-emission regions, the measured spacing was approximately 24~mm along both the $X$ and $Y$ directions. The two measurement planes were separated by approximately 16~mm along the optical axis. These results indicate that the measured scan region depended on the axial position of the measurement plane.

\section{Discussion}
\subsection{Focal-length-dependent plasma-voxel shape}
As the focal length increased, the plasma length increased from 1.39 to 4.64~mm, whereas the plasma width decreased overall. These changes indicate that a longer focal length produced a plasma voxel that was more elongated along the optical axis.

For an ideal Gaussian beam, the beam-waist radius is given by $w_0 = 2\lambda f/(\pi D)$, and the Rayleigh length is given by $z_R = \pi w_0^2/\lambda$, where $\lambda$ is the central wavelength and $D$ is the $1/e^2$ intensity diameter of the beam incident on the focusing lens. Thus, the beam-waist radius is proportional to the focal length, whereas the Rayleigh length is proportional to the square of the focal length. In this study, the confocal parameter $2z_R$ was calculated using a central wavelength of $800~\mathrm{nm}$ and the manufacturer's nominal beam diameter of $11~\mathrm{mm}$ and was used as a theoretical reference for comparison with the measured plasma length. The plasma length also increased with focal length and was therefore qualitatively consistent with the increase in the axial scale of the focal region. However, the measured plasma length exceeded the calculated confocal parameter $2z_R$ under all conditions. At $f = 75~\mathrm{mm}$ and $300~\mathrm{mm}$, the measured plasma lengths were $1.39~\mathrm{mm}$ and $4.64~\mathrm{mm}$, whereas the calculated $2z_R$ values were $0.10~\mathrm{mm}$ and $1.52~\mathrm{mm}$, respectively.

A possible instrumental contribution to the observed plasma elongation is spherical aberration in the focusing optics. However, for a fixed beam diameter, spherical aberration is expected to decrease with increasing focal length, whereas the measured plasma length increased with focal length. This opposite dependence suggests that spherical aberration is unlikely to be the dominant cause of the observed plasma elongation.

This difference indicates that the observed plasma-emission length cannot be described solely by ideal linear Gaussian focusing. During propagation of a high-intensity femtosecond laser pulse in air, Kerr self-focusing, photoionization, and defocusing by the generated plasma compete with one another, and a plasma channel can extend beyond the geometrical focal region \cite{Braun1995,Couairon2007,Berge2007,Kasparian2008,Chin2012}. Moreover, the strength of external focusing strongly affects plasma density and filament diameter, and a transition from conventional filamentation to localized optical breakdown has been reported under tight-focusing conditions \cite{Theberge2006,Liu2010}. Therefore, the emission region observed in this study may also have extended beyond the ideal depth of focus because of nonlinear propagation and optical-breakdown processes. It should be noted, however, that the emission region recorded by the camera is not identical to the electron-density distribution; rather, it is an apparent region influenced by emission intensity, scattered light, and imaging conditions \cite{Chen2010}. Furthermore, the nominal beam diameter used in the calculation did not account for the actual beam quality or wavefront aberrations. The calculated $2z_R$ should therefore be interpreted as a theoretical reference rather than as a direct prediction of plasma length.

The plasma length increased gradually from 75 to $150~\mathrm{mm}$, followed by a pronounced increase between 150 and $200~\mathrm{mm}$ and a further increase between 200 and $300~\mathrm{mm}$. While the calculated confocal parameter increased by a factor of 16 between $f = 75$ and $300~\mathrm{mm}$, the measured mean plasma length increased by a factor of approximately 3.35. This nonuniform dependence on focal length cannot be explained by a simple $f^2$ scaling and suggests that the focal intensity, ionization conditions, external focusing strength, and nonlinear propagation regime varied with focal length \cite{Theberge2006,Liu2010}. Because neither the intensity distribution during propagation nor the electron density was measured directly, the underlying mechanism cannot be identified from the present measurements alone.

From the perspective of display-system design, an ideal geometrical-optics model predicts a larger lateral focal displacement for a longer focal length at a fixed scan angle. The present measurements, however, show that increasing the focal length also increases the axial elongation of the plasma voxel while decreasing its width overall. Thus, focal length is a design parameter associated with a trade-off between potential display range and plasma-voxel shape. Because the scan range was not measured for each focal length in this study, this relationship requires further experimental verification.

\subsection{Scan-angle-dependent projected inclination of the plasma principal axis}

The projected inclination of the plasma principal axis varied approximately linearly with the galvanometer scan angle corresponding to each imaging plane. Particularly high linearity was obtained in the $Y$ direction, and the $X$ direction also showed an overall change in inclination with increasing scan angle. These results indicate that galvanometer scanning changes not only the generation position of a plasma voxel but also its projected orientation.

The response coefficients differed between the $X$ and $Y$ directions, and an offset in the projected inclination was observed even at zero scan angle. In a two-axis galvanometer scanning system, the two mirrors are not located at the same optical position, and the beam incidence position and angle at the lens can differ between the two scan axes; these factors can produce axis-dependent scan distortion \cite{Verboven1988}. The scan angles used in this study were free-space calibrated deflection angles obtained from beam displacement on a screen and did not directly represent the actual incidence conditions at the focusing lens.

At a nominal scan angle of $\pm 7^\circ$ and a focal length of 100~mm, the measured separation of laser irradiation marks increased from approximately 24 to 28~mm along both scanning axes between two measurement planes separated by approximately 16~mm along the optical axis. The plane at which a separation of 28~mm was measured corresponded to the plasma-generation position of the on-axis beam, whereas the plane at which a separation of 24~mm was measured corresponded to the upstream end of the off-axis plasma-emission regions. Assuming symmetric beam propagation, the observed 4~mm increase corresponds to an effective half-angle of approximately $7.13^\circ$, close to the nominal scan angle of $7^\circ$. This result is consistent with the geometrical prediction for the change in beam separation with axial position and suggests that the discrepancy between the measured 28~mm separation and the ideal focal-plane prediction of approximately 24.6~mm can be largely explained by the axial position of the measurement plane. However, this effective angle does not directly establish the actual beam incidence angle at the focusing lens. The causes of the sensitivity difference and offset between the axes therefore cannot be separated using the present measurements alone.

Across all 25 conditions, each projected angle depended mainly on the corresponding scan axis, although variation caused by the orthogonal axis was also observed. Thus, the voxel orientation during two-axis scanning may not be determined independently by the two scan angles. These results suggest that the voxel orientation is position dependent within the display region and may affect the apparent shape of rendered graphics.

\subsection{Influence of conical emission on display observation}

Figure~4 shows a graphic rendered by scanning plasma voxels together with the overall optical system during operation. During high-intensity femtosecond-laser-induced plasma generation, conically spreading light was observed along the optical axis in addition to the localized plasma emission. Strongly focused femtosecond laser pulses in air can generate supercontinuum and conical emission through nonlinear spectral broadening, and these emissions have been reported to appear as directional jets or complex color distributions depending on the conditions \cite{Xu2014,Ooi2016}. In the present experiment, a rainbow-colored pattern was also observed on a screen placed downstream.

Consequently, when plasma voxels are scanned to render a graphic, the observer may see not only the intended aerial plasma image but also the conical emission extending along the optical axis. This unwanted light differs from the intended plasma image and may reduce the visibility and contrast of the display. Therefore, in addition to controlling plasma-voxel shape and scanning characteristics, suppressing or blocking conical emission is an important issue in the use of femtosecond-laser-induced plasma for volumetric displays.
\begin{figure}
    \centering
    \includegraphics[width=1\linewidth]{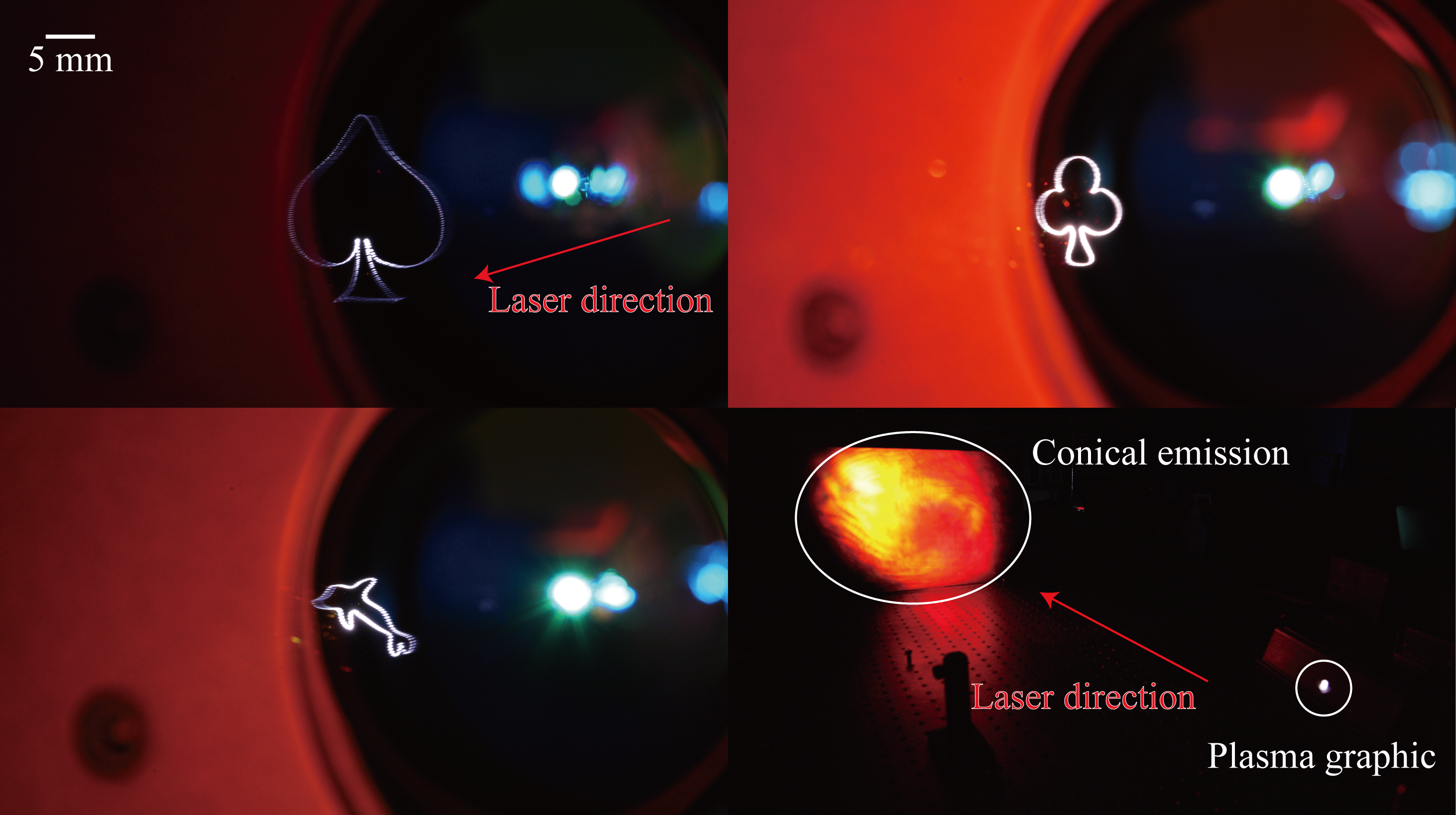}
    \caption{Examples of plasma-drawn patterns and the conical emission observed during plasma drawing at $f=75~\mathrm{mm}$. The scale bar applies to all plasma graphic images.}.
    \label{fig:placeholder}
\end{figure}
\section{Conclusion}

This study quantitatively evaluated the effects of focal length and scan angle on the shape and orientation of plasma voxels in a femtosecond-laser plasma volumetric display using a two-axis galvanometer mirror and a single focusing lens. As the focal length increased, the plasma length increased, while the plasma width decreased substantially at shorter focal lengths and changed only slightly at longer focal lengths. Consequently, the plasma voxels became increasingly elongated along the optical axis. The projected inclination of the plasma principal axis varied approximately linearly with scan angle.

Although volumetric displays based on femtosecond-laser-induced plasma have been reported previously, the present study directly related focal length and two-axis scan angle to the measured shape and orientation of plasma voxels in a basic single-lens configuration and quantitatively characterized these dependencies. The results suggest that similar evaluations could be used to predict and control the effects of other optical parameters on plasma voxels.

Future work will focus on correcting voxel shape and orientation as functions of scan position and on suppressing or blocking the supercontinuum and conical emission that propagate along the optical axis during plasma generation, thereby improving the visibility of the volumetric display.

\begin{backmatter}
\bmsection{Funding}
Pixie Dust Technologies, Inc.

\bmsection{Acknowledgments}
Codex (GPT-5.6 sol) was used to assist with the development and debugging of Python measurement scripts used in this study. All code, analysis procedures, and resulting outputs were reviewed and verified by the authors.

\bmsection{Disclosures}
YO: Pixie Dust Technologies, Inc. (F, E, I).

\bmsection{Data availability}
Data underlying the results presented in this paper are not publicly available at this time but may be obtained from the authors upon reasonable request.
\end{backmatter}

\bibliography{manuscript_references}

\end{document}


\maketitle

\section{Calibration from screen-position measurements}

The relationship between the galvanometer command value and the free-space beam-deflection angle was calibrated from laser-spot displacements on a screen. For a fixed command value, the transverse position of a beam propagating along a straight path can be expressed as
\begin{equation}
 p(z)=b+z\tan\theta,
 \label{eq:path}
\end{equation}
where $z$ is the screen position along the reference optical axis, $p(z)$ is the transverse spot position, $b$ is a position offset, and $\theta$ is the projected beam-deflection angle in the corresponding scanning plane. The angle can therefore be determined from measurements at two screen positions:
\begin{equation}
 \theta=\tan^{-1}\!\left[\frac{p(z_2)-p(z_1)}{z_2-z_1}\right].
 \label{eq:angle}
\end{equation}
Using the difference between two spot positions eliminates the constant offset $b$.

For a command increment of 5000 relative to the reference setting, the recorded transverse spot positions at $z_1=300\,\mathrm{mm}$ and $z_2=800\,\mathrm{mm}$ were 17 and 42\,mm for the X direction, and 18 and 44\,mm for the Y direction. Substituting these measurements into \eqref{eq:angle} gave $\theta_x=\tan^{-1}[(42-17)/(800-300)]\approx2.86^\circ$ and $\theta_y=\tan^{-1}[(44-18)/(800-300)]\approx2.98^\circ$.

\begin{table}[htbp]
 \begin{tabular}{ccccc}
 \toprule
 Scan axis & Command increment & $p(300\,\mathrm{mm})$ & $p(800\,\mathrm{mm})$ & Deflection angle \\
 \midrule
 X & 5000 & 17\,mm & 42\,mm & $2.86^\circ$ \\
 Y & 5000 & 18\,mm & 44\,mm & $2.98^\circ$ \\
 \bottomrule
 \end{tabular}
 \centering
 \caption{Screen-position measurements used to determine the angular conversion coefficients.}
 \label{tab:calibration}
 \small
\end{table}

These coefficients were incorporated into the control program to convert angular commands into galvanometer command values. For the axis-specific angular commands $\theta_{x,\mathrm{cmd}}$ and $\theta_{y,\mathrm{cmd}}$, expressed in degrees, the command values were calculated as $u_x=u_{x,0}+(5000/2.86)\theta_{x,\mathrm{cmd}}$ and $u_y=u_{y,0}+(5000/2.98)\theta_{y,\mathrm{cmd}}$, followed by rounding to integers. In the control implementation, these angular commands were obtained after the programmed scan-axis tilt correction.

The angles determined here describe the free-space beam deflection. They are not measurements of the mechanical mirror rotation or direct measurements of the incidence angle at the focusing lens. The numerical precision of the stored conversion coefficients does not represent an experimentally established angular uncertainty.

\section{Spot displacement as a function of screen distance}

A single acrylic screen coated with a material that emits light in response to infrared irradiation was used. The screen was first positioned at $z=100\,\mathrm{mm}$ along the reference optical axis and imaged. The same screen was then translated away from the source along the reference optical axis in increments of 150\,mm, and images were acquired sequentially at $z=250$, 400, 550, and 700\,mm.

At each position, the screen remained stationary while the galvanometer commands were changed to record the reference spot and spots displaced rightward, leftward, upward, and downward in the image. The same command settings were used at all five screen positions, giving 25 images in total.

Four ArUco markers attached to the screen defined a common image coordinate system. The marker centers were mapped to the corners of a $1000\times700$-pixel image using a planar projective transformation. This transformation provided a consistent screen-coordinate representation across the five axial positions.

The infrared laser irradiation position was recorded as a spot produced by emission from the coating. Intensity thresholding and connected-component analysis were applied to the acquired images to extract the spot region, and the centroid of the selected binary region was used as the irradiation position.

For each screen position, the reference-spot centroid was subtracted from the centroid measured at each displaced setting:
\begin{equation}
 \Delta p_{j,k}=p_{j,k}-p_{0,k},
 \label{eq:relative}
\end{equation}
where $j$ denotes the displaced setting, $k$ denotes the screen position, and $p_{0,k}$ is the corresponding reference-spot coordinate. Horizontal coordinates were used for the rightward and leftward settings, and vertical coordinates for the upward and downward settings.

The resulting displacements were fitted independently using
\begin{equation}
 \Delta p_j(z)=a_jz+c_j.
 \label{eq:fit}
\end{equation}
The intercept $c_j$ was fitted freely rather than constrained to zero. Displacements were retained in pixels for this linearity analysis; conversion to physical length is unnecessary for evaluating the coefficient of determination under a constant image scale.

\begin{table}[htbp]
 \begin{tabular}{lcc}
 \toprule
 Direction in the source image & Number of screen positions & $R^2$ \\
 \midrule
 Rightward & 5 & >0.99 \\
 Leftward & 5 & >0.99 \\
 Upward & 5 & >0.99 \\
 Downward & 5 & >0.99 \\
 \bottomrule
 \end{tabular}
 \centering
 \caption{Linearity of the reference-subtracted spot displacement over the five screen positions.}
 \label{tab:linearity}
 \small
\end{table}

All four displacement series showed an approximately linear dependence on screen position, with $R^2>0.99$. These results support the use of a straight-line propagation model for relating spot displacement to screen distance at the tested fixed command settings.

The two-distance measurements in Section~1 establish the angular conversion coefficients, whereas the five-position image measurements characterize the linearity of spot displacement with axial distance. The latter measurements do not independently establish the absolute angular scale or the linearity of angle versus command value over the full command range.

\begin{figure}
    \centering
    \includegraphics[width=1.0\linewidth]{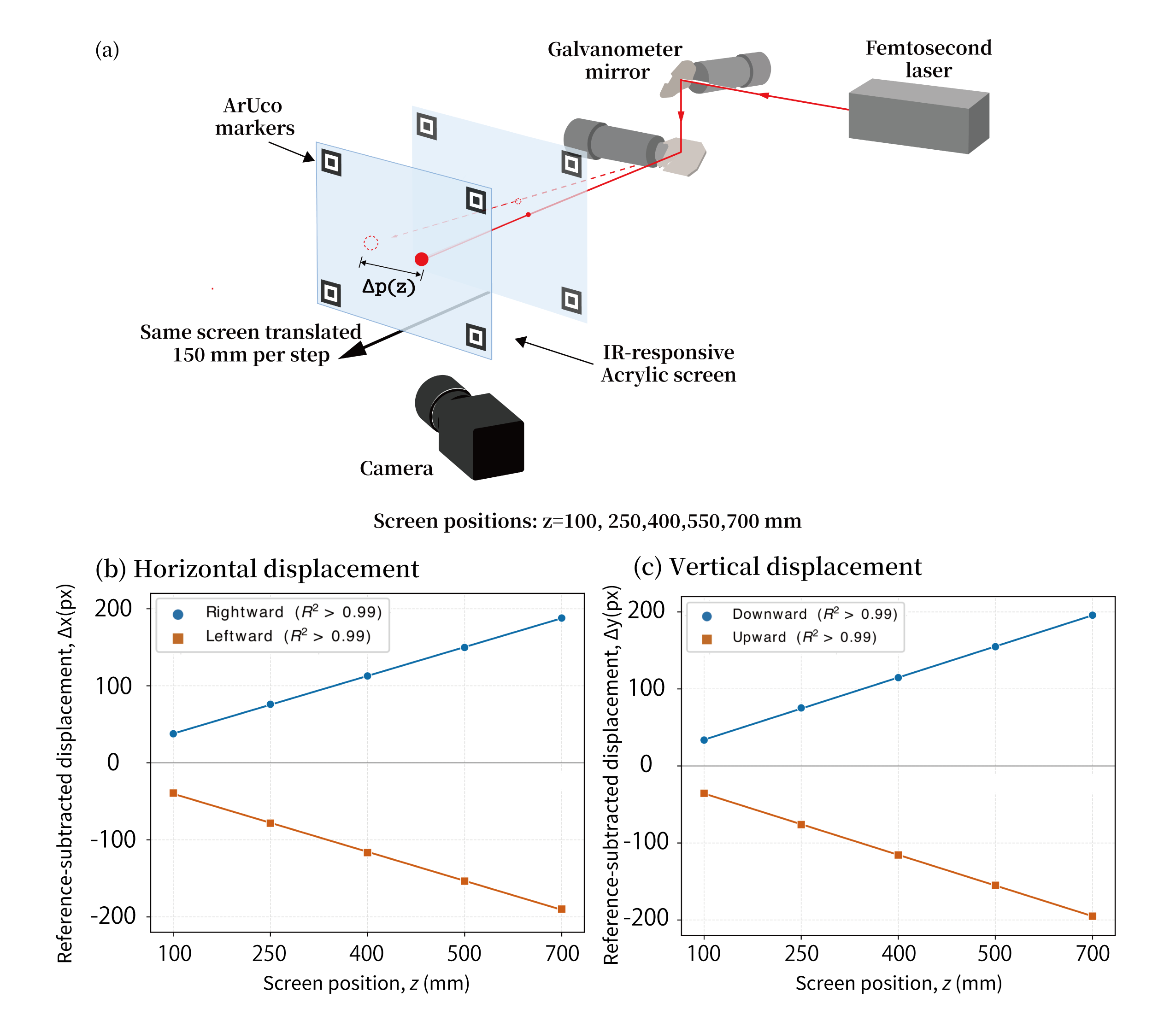}
    \caption{Calibration and distance dependence of galvanometer-deflected laser spots.
(a) Schematic of the measurement arrangement. A single acrylic screen coated with an infrared-responsive luminescent material was translated away from the source along the reference optical axis in 150\,mm increments, and spot images were acquired at each position. The red line represents the deflected beam, and the dashed line represents the reference optical axis. The red and open points on the screen represent the displaced and reference positions recorded at different command settings, respectively. (b) Reference-subtracted horizontal spot displacements for the rightward and leftward settings at $z=100$, 250, 400, 550, and 700\,mm. (c) Corresponding vertical spot displacements for the upward and downward settings. Symbols represent centroid measurements, and solid lines represent independent linear fits with freely fitted intercepts. Image coordinates increase rightward and downward. Panels (b) and (c) use pixel units and evaluate displacement linearity; the angular conversion coefficients were obtained from the physical measurements summarized in Table~\ref{tab:calibration}.}
    \label{fig:placeholder}
\end{figure}

\section{Effect of relative intensity threshold on plasma-width measurement}

The influence of the relative intensity threshold on plasma-width measurement was evaluated at $f = 75$ and $100~\mathrm{mm}$. Figure~S2 compares the extracted plasma-emission contours and width measurement positions obtained using thresholds of 35\% and 45\% of the peak intensity value within the reference region at $f = 75~\mathrm{mm}$.
At a threshold of 35\%, the surrounding diffuse glow affected the extracted emission region, resulting in an increased measured width at $f = 75~\mathrm{mm}$. Increasing the threshold to 45\% reduced the influence of the diffuse glow on the extracted contour. The mean measured plasma width decreased from $0.411 \pm 0.016~\mathrm{mm}$ at 35\% to $0.360 \pm 0.015~\mathrm{mm}$ at 45\%, corresponding to a reduction of 12.4\%. For comparison, the mean measured width at $f = 100~\mathrm{mm}$ decreased from $0.333 \pm 0.004~\mathrm{mm}$ to $0.313 \pm 0.005~\mathrm{mm}$, corresponding to a reduction of 5.9\%. The uncertainties represent one standard deviation from ten images at each focal length.
A relative threshold of 35\% was used as the standard criterion for plasma-width measurement across the focal-length conditions. However, at $f = 75~\mathrm{mm}$, a threshold of 45\% was adopted to reduce the influence of the surrounding diffuse glow on the extracted plasma contour.

\begin{figure}
    \centering
    \includegraphics[width=1.0\linewidth]{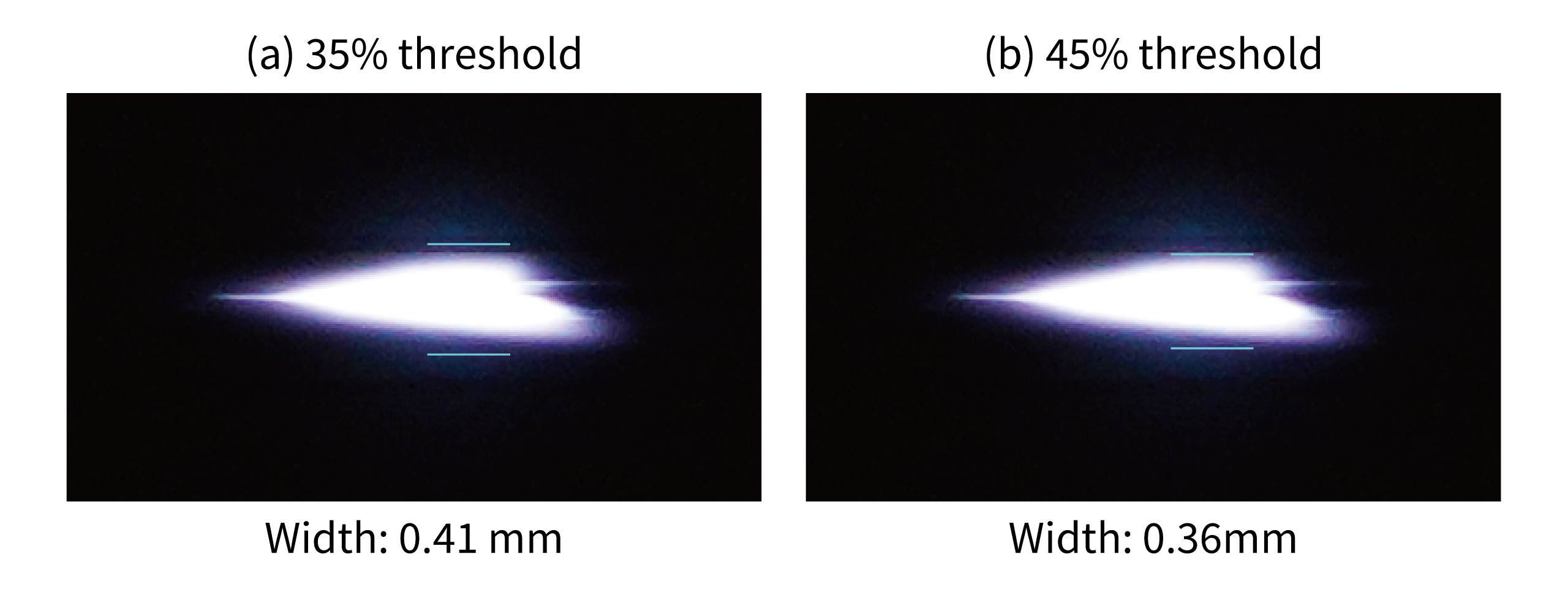}
    \caption{Comparison of plasma-width measurements at $f = 75~\mathrm{mm}$ using relative intensity thresholds of (a) 35\% and (b) 45\%. The extracted emission contours and width measurement positions are overlaid on the same side-view JPEG image.}
    \label{fig:placeholder}
\end{figure}
